\documentclass[12pt,preprint]{aastex} 

\usepackage{epsfig}
\usepackage{verbatim}
\usepackage{graphicx}
\usepackage{amssymb}
\usepackage{natbib}
\newcommand\flu{$\mathrm{erg}~\mathrm{cm}^{-2}$}
\newcommand\flux{$\mathrm{erg}~\mathrm{cm}^{-2}~\mathrm{s}^{-1}$}
\newcommand\nuflux{$\mathrm{erg}~\mathrm{cm}^{-2}~\mathrm{s}^{-1}~\mathrm{\AA}^{-1}$}
\newcommand\cps{$\mathrm{counts}~\mathrm{s}^{-1}$}
\def\lesssim{\mathrel{\hbox{\rlap{\hbox{\lower4pt\hbox{$\sim$}}}\hbox{$<$}}}}
\def\gtrsim{\mathrel{\hbox{\rlap{\hbox{\lower4pt\hbox{$\sim$}}}\hbox{$>$}}}}

\begin{document} 

\title{Swift-UVOT Observations of the X-Ray Flash 050406}
\author{P.~Schady\altaffilmark{1,2}, K.O.~Mason\altaffilmark{3}, J.P.~Osborne\altaffilmark{4}, M.J.~Page\altaffilmark{2}, P.W.A.~Roming\altaffilmark{1}, M.~Still\altaffilmark{5,6}, B.~Zhang\altaffilmark{7}, A.J.~Blustin\altaffilmark{2}, P.~Boyd\altaffilmark{5}, A.~Cucchiara\altaffilmark{1}, N.~Gehrels\altaffilmark{5}, C.~Gronwall\altaffilmark{1}, M.~De~Pasquale\altaffilmark{2}, S.T.~Holland\altaffilmark{5}, F.E.~Marshall\altaffilmark{5}, K.E.~McGowan\altaffilmark{2}, J.A.~Nousek\altaffilmark{1}}

\altaffiltext{1}{Department of Astronomy and Astrophysics, Pennsylvania State University, 525 Davey Laboratory, University Park, PA 16802, USA.}
\altaffiltext{2}{The UCL Mullard Space Science Laboratory, Holmbury St Mary, Dorking, Surrey, RH5 6NT, UK.}
\altaffiltext{3}{The Particle Physics and Astronomy Research Council, Polaris House, North Star Avenue, Swindon, Wiltshire, SN2 1SZ, UK.}
\altaffiltext{4}{X-ray and Observational Astronomy Group, Department of Physics \& Astronomy, University of Leicester, LE1 7RH, UK.}
\altaffiltext{5}{NASA/Goddard Space Flight Center, Greenbelt, MD 20771, USA.}
\altaffiltext{6}{South African Astronomical Observatory, PO Box 9, Observatory 7935, South Africa}
\altaffiltext{7}{Department of Physics, University of Nevada, Las Vegas, NV 89154, USA.}


\begin{abstract}
We present {\it Swift}-UVOT data on the optical afterglow of the X-ray flash of 2005 April 6 (XRF~050406) from 88~s to $\sim 10^5$~s after the initial prompt $\gamma$-ray emission. Our observations in the {\it V}, {\it B} and {\it U} bands are the earliest that have been taken of an XRF optical counterpart. Combining the early-time optical temporal and spectral properties with $\gamma$- and simultaneous X-ray data taken with the BAT and XRT telescopes on-board {\it Swift}, we are able to constrain possible origins of the XRF. The prompt emission had a FRED profile (fast-rise, exponential decay) with a duration of $\mathrm{T}_{90}=5.7~\mathrm{s}\pm 0.2~\mathrm{s}$, putting it at the short end of the long-burst duration distribution. The absence of photoelectric absorption red-ward of $4000$~\AA ~in the UV/optical spectrum provides a firm upper limit of $\mathrm{z}\leq 3.1$ on the redshift, thus excluding a high redshift as the sole reason for the soft spectrum. The optical light curve is consistent with a power-law decay with slope $\alpha = -0.75\pm 0.26~(F_{\nu}\propto t^{\alpha})$, and a maximum occurring in the first 200~s after the initial $\gamma$-ray emission. The softness of the prompt emission is well described by an off-axis structured jet model, which is able to account for the early peak flux and shallow decay observed in the optical and X-ray bands.
\end{abstract} 


\section{Introduction}\label{sec:intr}
X-ray flashes (XRFs) are a softer class of gamma-ray bursts (GRBs) with a $\nu F_{\nu}$ spectrum that peaks at energies $E_{\mathrm{p}} \lesssim 30$~keV~\citep{hik+01}. They make up around a third of the GRB population and their defining characteristic is a high X-ray fluence ($S_{\mathrm{X}}$) relative to their gamma-ray fluence ($S_{\gamma}$) such that log[$S_{\mathrm(X)}$/$S_{\gamma}$]$ >0$. Classical GRBs have log[$S_{\mathrm{X}}$/$S_{\gamma}$]$ <-0.5$, and bursts with intermediate fluence ratios are referred to as X-ray rich (XRR) bursts.

All three classes of burst are distributed isotropically on the sky and share many temporal and spectral properties. Their prompt emission light curves are similar in duration and structure, and their spectra are well fit by the `Band' function~\citep{bmf+93}. The differentiating feature in the context of the Band model is the spectral break energy. Most GRBs have observed peak energies in the range $E_{\mathrm{p}}\sim 135-350$~keV~\citep{pbm+00}, whereas XRFs have $E_{\mathrm{p}}\ll 100$~keV that can be as low as 3~keV~\citep{kwh+01,bol+03,slk+05}.

Other similarities exist in a number of empirical laws that are satisfied by all three classes of bursts. The burst populations form a continuum in the [$S(2-400$~keV), $E_{\mathrm{p}}$]-plane~\citep{stk+03}, and the Amati relation~\citep{aft+02} is satisfied by at least three XRFs~\citep{ldg05,sbb+05}, which relates a GRB's peak energy $E_{\mathrm{p}}$ to its isotropic equivalent energy $E_{\mathrm{iso}}$ such that $E_{\mathrm{p}}\propto E_{\mathrm{iso}}^{1/2}$~\citep{ldg05}. 

Theoretical models that aim to explain the origins of XRFs include (1) GRBs at high redshift~\citep{hei03}, which in this context is at redshifts $\mathrm{z} > 5$, (2) a baryon loaded fireball (dirty fireball model)~\citep{dcb99}, (3) a clean fireball ($\Gamma\gg 300$) where the difference in Lorentz factor between colliding shells is small~\citep{zm02,bdm+05}, (4) a photosphere-dominated model~\citep{rl02,mrr+02}, and (5) classical GRBs seen off-axis, either within the context of a uniform jet model~\citep{yin02} or a universal structured jet model~\citep{zdl+04}.

The broad range of properties shared by XRFs, XRR bursts and GRBs suggests that all three classes are due to the same underlying phenomenon and that there should be a single model to explain their origins. \citet{rl02} and \citet{mrr+02} suggested a dominating baryonic photosphere with an extended optically thick region caused by pair-producing shocks as a source of excess X-ray emission. However, such a model requires additional considerations to account for those bursts with $E_{\mathrm{p}} = 3-5$~keV, as is the case for XRF~020427~\citep{acf+02}, XRF~020903~\citep{skb+04} and XRF~010213~\citep{ssk+04}. Models that require the GRB jet to be observed off-axis have been particularly successful at explaining observations of XRFs (e.g. Fynbo et al. 2004; Soderberg et al. 2005), and their ability to describe both GRBs and XRFs in a unified way have made them increasingly popular.

Afterglows have now been detected for a number of XRFs from the X-ray down to radio wavebands~\citep{lg03}. The first X-ray and radio afterglows were discovered in 2001 (XRF~011030; Taylor et al. 2001; Harrison et al. 2001), and in 2002 the first XRF optical afterglow was observed (XRF~020903; Soderberg et al. 2004). Since then only a handful of XRFs have had associated optical counterparts (e.g. XRF~030429, XRF~030528, XRF~030723, XRF~040916, XRF~050315, XRF~050416a), with the earliest observation of an optical afterglow detection taking place 31.3~mins after the initial burst (XRF~030723; Smith et al. 2003).

In this paper we describe {\it Swift} observations of XRF~050406, focusing in particular on the optical/UV band data (see~\citet{rmb+05} for a detailed analysis of the X-ray data). These are the earliest observations taken of an XRF optical counterpart to date, beginning only 88~s after the initial emission. In section~\ref{sec:obs} we review the observations made by all 3 instruments on-board {\it Swift} followed by an analysis of the temporal and spectral behaviour of the XRF in the UV, optical and X-ray energy bands in sections~\ref{sec:lc} and \ref{sec:spec}. In section~\ref{sec:disc} we discuss the results and investigate their implications for XRF theoretical models. The conclusions are summarised in section~\ref{sec:concs}.

\section{Observations}\label{sec:obs}
On 2005 April 6 the Swift Burst Alert Telescope (BAT; Barthelmy et al. 2005) triggered and located on-board XRF~050406, at 15:58:48 UT~\citep{pbc+05}. The BAT location calculated on-board was RA =$02^{\mathrm{h}} 17^{\mathrm{m}} 42.7^{\mathrm{s}}$, Dec=$-50^{o}10\arcmin 40.8\arcsec$ (J2000; Parsons el al. 2005). This was later refined to RA=$02^{\mathrm{h}} 17^{\mathrm{m}} 53^{\mathrm{s}}$, Dec=$-50^{o}10\arcmin 51.6\arcsec$ with an uncertainty of $3\arcmin$ (95\% containment) using the full dataset downloaded through the Malindi ground-station $\sim 1$ hour after the burst~\citep{kbb+05}. This corresponds to a Galactic latitude of $-61.7^{o}$ with a local reddening of $E(\bv )$ = 0.022 mag~\citep{sfd98} and a Galactic column density of $N_{H} = 2.8\times 10^{20}~\mathrm{cm}^{-2}$~\citep{dl90}.

The light curve comprised of a single peak showing a Fast-Rise and Exponential Decay (FRED) profile with a $\mathrm{T}_{90}$ burst duration of $5.7\pm 0.2$~s in the 15-350~keV band~\citep{kbb+05}. The burst was very soft, with most flux emitted below 50~keV. The time-averaged spectrum was well fit by a simple power-law model ($\chi ^{2}=94$ for 73 d.o.f.), a cutoff power-law model ($\chi ^{2}=88$ for 72 d.o.f.), and the Band spectral model ($\chi ^{2}=89$ for 71 d.o.f.). The best fit $E_{\mathrm{p}}$ value in the latter two models was $E_{\mathrm{p},\mathrm{cutoff}}=15.3^{+34.9}_{-6.2}$~keV and $E_{\mathrm{p},\mathrm{Band}}=24.5^{+35.3}_{-12.7}$~keV. A simple power-law spectral fit does not constrain the value of $E_{\mathrm{p}}$. However, the steep best-fit spectral index of $\beta = -2.56^{+0.39}_{-0.33}$ suggests that the observed $\gamma$-ray spectrum corresponds to the high-energy portion of the Band model. In this case a spectral break must have occurred near or below the lower limit of the BAT energy band, at $\sim 15$~keV. The $T_{90}$ fluence in the 15-350~keV band was $1.0\times 10^{-7}$~\flu, with $6.15\times 10^{-8}$~\flu ~lying in the 15-50~keV band, and $8.2\times 10^{-8}$~\flu ~in the 15-100~keV band, making up more than $80\%$ of the total fluence.

The spacecraft executed an automated slew to the BAT position, and within 72~s the X-Ray Telescope (XRT; Burrows et al. 2005a) was on target and imaging the field. Ground analysis of the data revealed a fading X-ray source within the BAT error circle $28\arcsec$ from the BAT position~\citep{cpr+05}.\par

The Ultra-Violet/Optical Telescope (UVOT; Roming et al. 2005a) began settled observations of the field of XRF~050406 88~s after the BAT trigger, and carried out an automated series of observations amounting to 89 exposures in 3 optical and 3 ultra-violet filters. The first exposure taken was a 100~s {\it V}-filter `finding chart' followed by 10~s exposures in each passband for 9 rotations of the filter wheel, after which a series of $\sim 100$~s and then $\sim900$~s exposures were made.\par

\section{Results}
Initial analysis of the {\it V} band early time data processed on-board revealed no new source in the field down to a $3\sigma$ background limit of {\it V}$> 18.8$~\citep{lhb+05}. However, analysis of the {\it V} band image data, which is processed on the ground, revealed a faint source $3.3\sigma$ above background within the reported $5\arcsec$ radius XRT error circle, with $V = 19.44\pm 0.32$. The initial lack of detection was the result of the difference in the drift corrections applied on-board to that applied on the ground processed data. The lower accuracy of the on-board drift corrections caused the optical afterglow to be lost within the background noise in the first dataset analysed. The position of the uncatalogued source is RA = $02^{\mathrm{h}} 17^{\mathrm{m}} 52.2^{\mathrm{s}}$, Dec = $-50^{o}11\arcmin 15.5\arcsec$ (J2000) with an estimated uncertainty of $0.5\arcsec$, consistent with the position reported by~\citet{bog05} who detected a faint source in r and i band observations taken with the Magellan/Clay telescope. The source was no longer observed above background in any of the subsequent {\it V}-band images, either in the individual or co-added exposures.

\begin{figure}
\epsscale{1}
\plotone{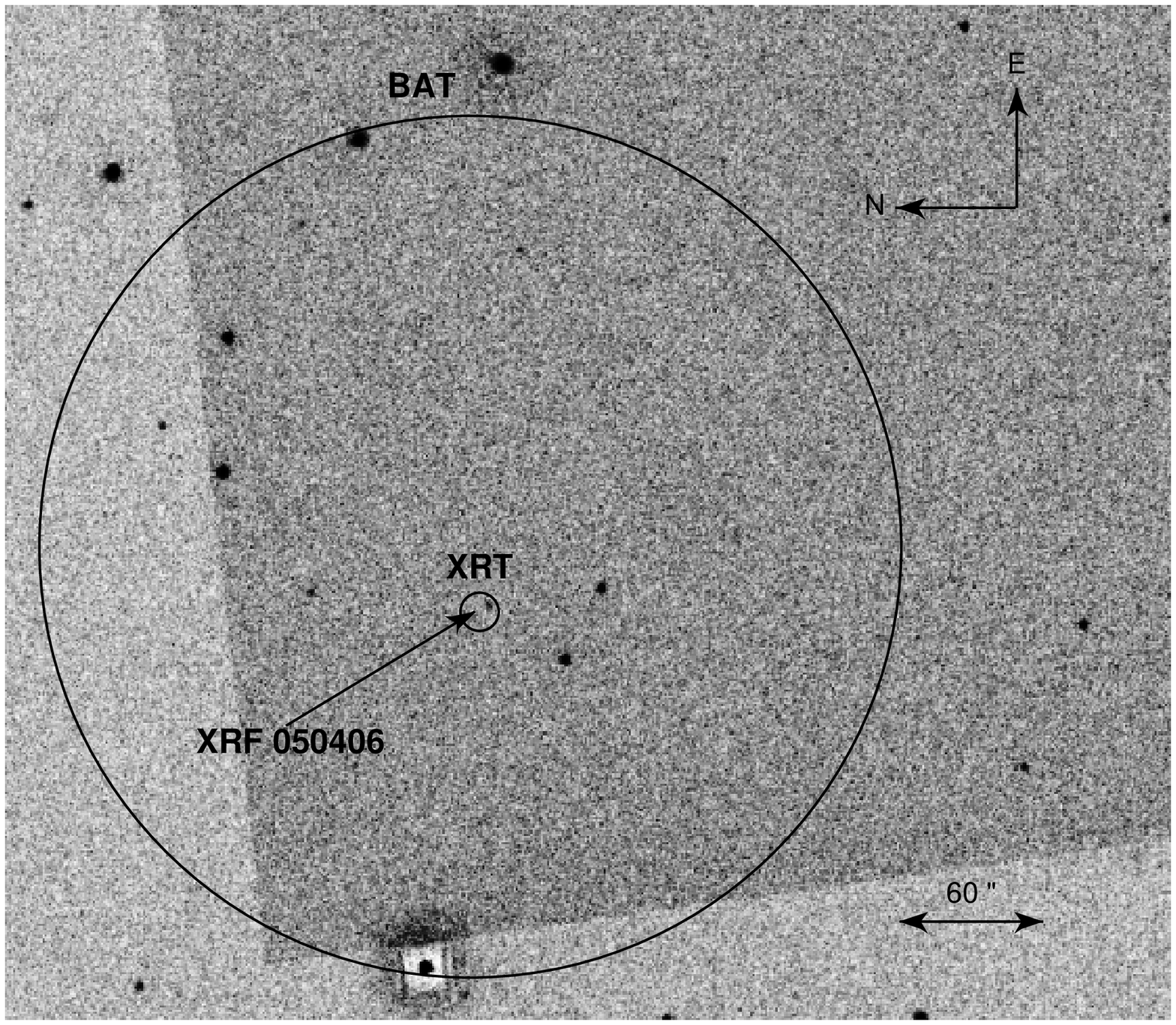} 
\caption{{\footnotesize Stacked UVOT image composed of 100~s {\it V} band exposure and several {\it V}, {\it B} and {\it U} band exposures from observation $<~1000$~s after the burst prompt emission. The transient source is at RA = $02^{\mathrm{h}} 17^{\mathrm{m}} 52.2^{\mathrm{s}}$, Dec = $-50^{o}11\arcmin 15.5\arcsec$ (J2000), with the $3\arcmin$ BAT error circle and $6\arcsec$ XRT error circle overlaid. The total exposure time of the stacked image is 354~s. The step change in background density across the image is a result of combining observations made with different data window sizes.}}\label{fig:UBVimage}
\end{figure}

The source was detected in the UVOT {\it B} and {\it U} bands, with $B = 19.30\pm 0.66$ and $U = 18.55\pm 0.59$ at 423~s and 226~s after the prompt emission, respectively. $1000$~s after the burst trigger the source was no longer detected above background in either the {\it V}, {\it B} or {\it U} band, confirming this source as the fading afterglow of XRF~050406. A co-added image (Fig.~\ref{fig:UBVimage}) constructed from all exposures in the {\it V}, {\it B} and {\it U} bands up to 1000~s after the burst provided a $8.7\sigma$ level detection.

\subsection{The Light Curve}\label{sec:lc}
Photometry of the images was carried out using a $2\arcsec$ radius aperture, which was found to maximise the signal-to-noise (S/N) ratio of the source detection. This is smaller than the PSF FWHM corresponding to each filter, and therefore a correction was applied to account for the loss of source photon counts in the wings of the PSF. The best-fit power law decay indices to the {\it B} and {\it U} light curves are poorly constrained due to the faintness of the source, with $\alpha_{B} = -0.74\pm 0.52$ and $\alpha_{U} = -1.19\pm 0.56$, and a combined fit to both filters yields $\alpha_{B,U} = -0.77\pm 0.67$ ($F_{\nu} \propto \mathrm{t}^{\alpha}$). All errors quoted in this section are at the $1\sigma$ level.

To obtain a higher S/N light curve the neighbouring pairs of {\it U} and {\it B} exposures were co-added, on the assumption that there was no colour evolution during this time period. It is possible that the cooling break could have passed through the optical bands at this time, causing a spectral break between the bands and a difference in their temporal decay rate. The quality of the data does not allow for this to be verified. However, such a migration of the cooling frequency is typically observed to occur at later times, on the order of $\sim 10^4$~s (e.g. see Blustin et al. 2005), thus making such an assumption reasonable. This provided a total of nine images with an effective exposure of $2\times 10$~s and two images with effective exposures of $2\times 900$~s, where the last two co-added exposures provided only upper limits. The co-added {\it UB} exposures yield a best-fit power-law decay index of $\alpha_{UB} = -0.75\pm 0.26$ over the time interval $\sim$T+200 to $\sim$T+6000~s. This is consistent with the {\it V} band observations where the afterglow must have decayed at a rate of $\alpha_{V}\sim -0.86$ to account for the drop in count rate from 0.22~\cps ~down to below the background level in $\sim 460$~s. The {\it V} band light curve consists of a 100~s exposure, a $8\times 10~$s exposure and a 900~s exposure, and is shown against the {\it UB} light curve in Figure~\ref{fig:lc}. The source magnitudes for the independent {\it V}, {\it B}, and {\it U} exposures and the combined {\it UB} exposures upto $\sim 10^{4}$~s are listed in table~\ref{tab:timephot}. Observations that have count rates lower than $< 0.1$~\cps are treated as upper limits, and it is the error on the background subtracted count rate that is given as the upper limit.

\begin{figure}
\epsscale{1}
\plotone{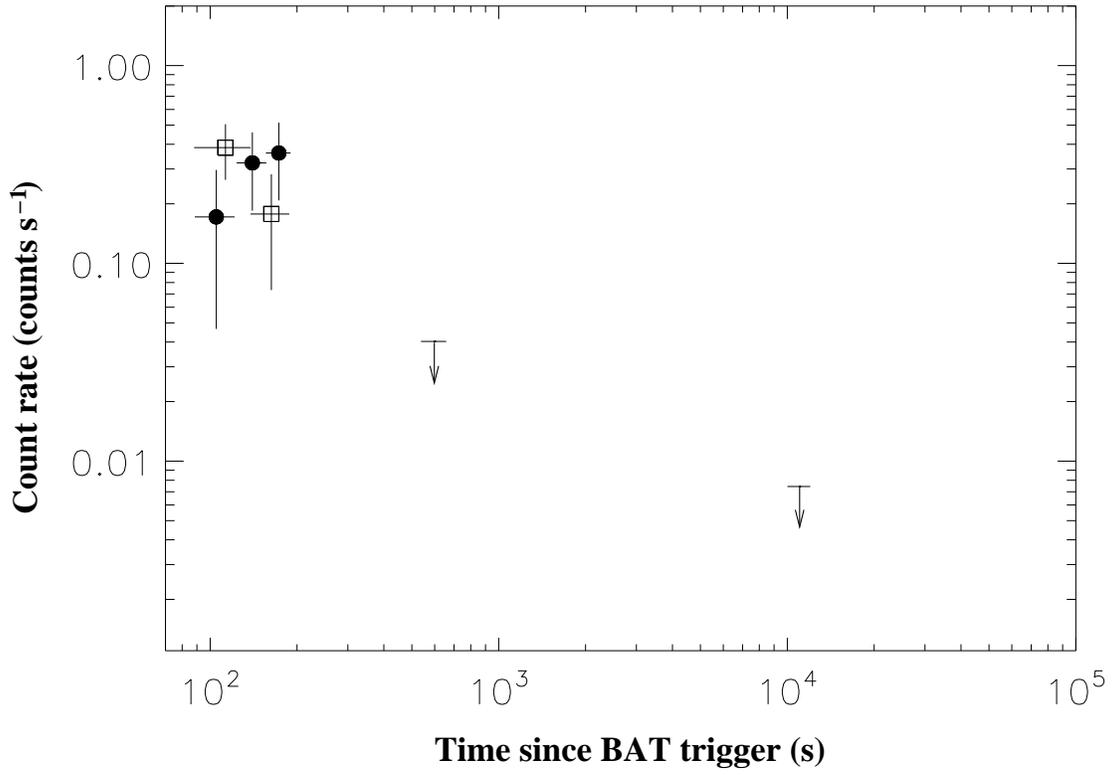}
\caption
{{\footnotesize {\it V} filter light curve where the first 100~s observation has been split up into $2\times 50$~s segments (open squares), and 3 segments made up of 35~s, 35~s and 30~s (filled circles). The light curve decays for a split into $2\times 50~$s and rises when split into three intervals of similar length.}}\label{fig:Vlc}
\end{figure}

\begin{figure}
\epsscale{1}
\plotone{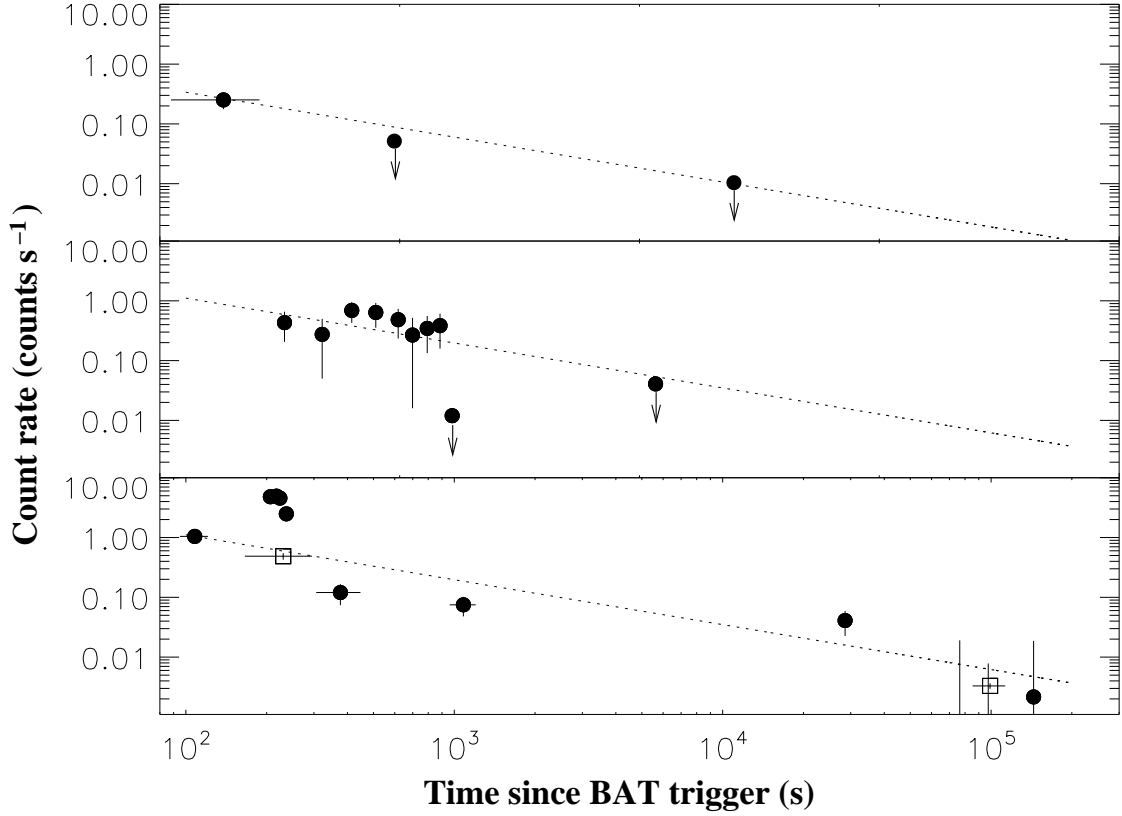}
\caption{{\footnotesize {\it V} light curve (top), co-added {\it U} and {\it B} light curve (middle), and X-ray light curve in 0.2-10~keV energy band (bottom) for XRF~050406. The X-ray light curve contains windowed timing (WT) and photon-counting (PC) mode data, shown as filled circles and empty squares respectively. The best power-law fit to the {\it UB} co-added data points is shown in each panel, and this has a decay index $\alpha = -0.75\pm 0.26$. The power-law decay curve is renormalized in the top panel to pass through the first {\it V} data point.}}\label{fig:lc}
\end{figure}

Event-mode data are available for all filters for the first 1200~s of UVOT observations, in which timing information is stored with a resolution of 11~ms. The source count rate in the {\it V} filter during the first 100~s of data was only $0.25\pm 0.08$~\cps, which restricts the temporal information available during this time interval. A Kolmogorov-Smirnov (K-S) test was performed to compare the temporal distribution of counts during the first 100~s {\it V} exposure (T+88 to T+188 sec) with that observed from a source decaying as a power-law with photon index $\alpha = -0.75$ (corresponding to the {\it UB} light curve decay rate). The probability that the model and observed dataset derive from the same distribution is $\mathrm{P}_{\mathrm{K-S}}=0.27$. For comparison, the observed dataset is consistent to constant emission with a probability $\mathrm{P}_{\mathrm{K-S}}=0.45$, and it has a probability $\mathrm{P}_{\mathrm{K-S}}=0.68$ of being consistent to a source brightening as a power law with photon index $\alpha=0.75$. To illustrate the difficulty associated with determining the optical afterglow temporal behaviour in the first 100s UVOT observations, Figure~\ref{fig:Vlc} shows the change observed in the {\it V} band light curve when the finding chart exposure is split up into segments of differing lengths. A rising behaviour is derived when the finding chart is split into three segments of 35~s, 35~s and 30~s exposures, whereas a $2\times 50~$s split produces a decaying light curve.

It is worth noting that during this time period a bright flare is observed in the X-ray light curve (Fig.~\ref{fig:lc}, bottom panel), where the light curve brightened by a factor of 6 between 150 and 213~s post-burst. Based on the best-fit decay slope to the co-added UB filters, which predicts a flux of $\sim 0.63$~\cps~ at 213~s, an equivalent re-brightening in the optical band during this time would have caused the flux to be 3.78~\cps; more than 8 times the observed count rate in the first UB bins.

\subsection{Spectrum}\label{sec:spec}
The multi-wavelength observations carried out by the UVOT provide spectral information in the wavelength range $1700$~\AA-$6000$~\AA. Using each optical and UV filter as an energy bin, a spectrum was produced for the time period 220-950~s after the BAT trigger. This interval is when the filters are cycling rapidly, ensuring that each band is sampling the same average in the decay curve. All quoted errors on the best-fit spectral parameters are at the 90\% confidence level.

The UVOT exposures taken during this time period were co-added according to filter, producing an $8\times 10$~s exposure for the {\it V}-band, and $9\times 10$~s exposures in all other filters. UVOT source and background data files compatible with the spectral fitting code XSPEC (v12.2.0; Arnaud 1996) were created using the tool `uvot2pha'. The source extraction region used was selected to optimise the S/N ratio. To maintain consistency throughout the filters the source aperture size was chosen to contain $\sim 75\%$ of the counts emitted by the afterglow, which varied between $2.0\arcsec$ and $2.9\arcsec$, and an aperture size correction was applied. The background was determined from an annular region around the source. The count rates and corresponding fluxes for each of the UVOT exposures used in the spectrum are listed in table~\ref{tab:specphot}.

A fit using a simple power law model that took into account Galactic reddening provided a statistically unacceptable fit ($\chi^{2} = 15.7$ for 4 d.o.f.), with best-fit photon index $\beta_{\mathrm{UVOT}} = -3.60$ ($F_{\nu}=\nu ^{\beta}$). The addition of a Lyman absorption edge at a free floating redshift significantly improved the fit ($\chi^{2} = 2.3$ for 2 d.o.f.), and yielded a spectral index of $\beta_{\mathrm{UVOT}} = -0.82^{+3.58}_{-2.18}$ and $\mathrm{z} = 2.70^{+0.29}_{-0.41}$.
 
The analysis was repeated employing Cash statistics~\citep{cas79}, which uses a likelihood function to fit a model to the data. This is more appropriate given the low source photon count rate and provides more rigorous confidence intervals. A re-fit to the spectrum using a power-law model with Galactic absorption and extinction included and a Lyman absorption edge at a free floating redshift produced best-fit parameters consistent with those found with a $\chi^2$ minimisation fit. The fit yielded $\beta_{\mathrm{UVOT}} = -0.94\pm 2.78$ and $\mathrm{z} = 2.50^{+0.43}_{-0.42}$.

\begin{figure}
\epsscale{1}
\plotone{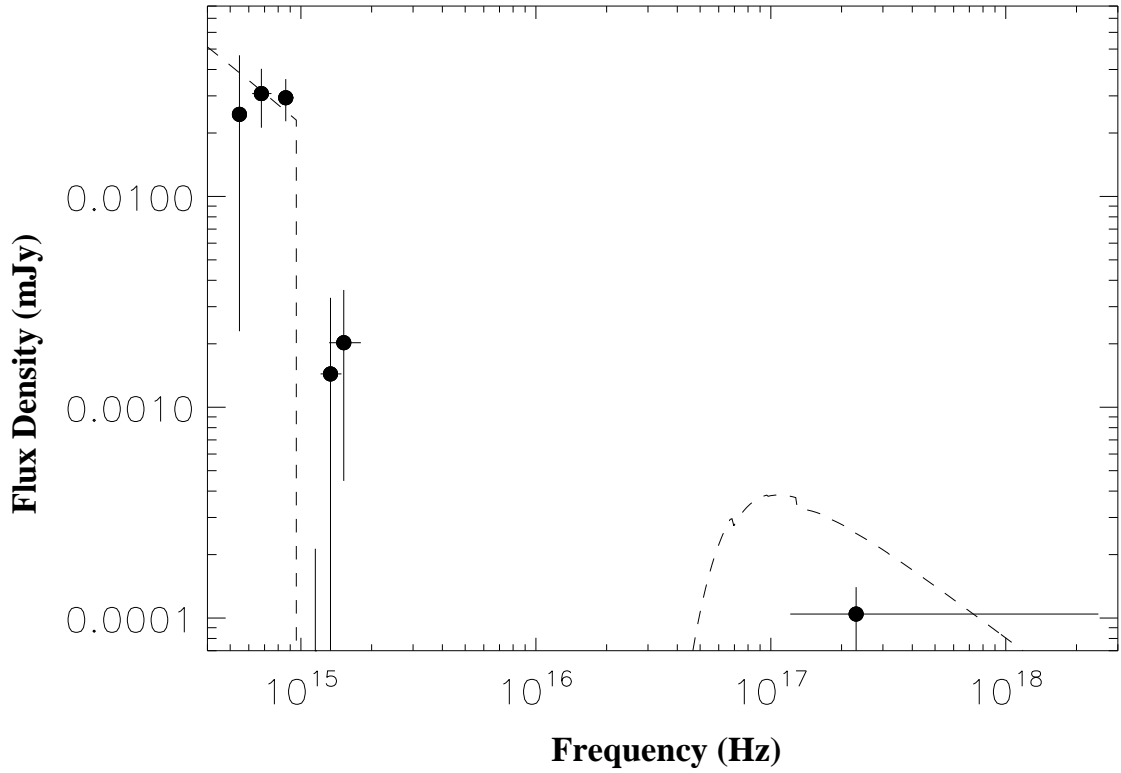} 
\caption{{\footnotesize Combined UVOT and X-ray spectrum of XRF~050406 between the epochs T+220s and T+950s, when the UVOT filter wheel was cycling rapidly. The dashed line represents the best-fit power-law model with reddening and absorption from Galactic material incorporated, as well as an additional redshifted Lyman absorption edge.}}\label{fig:spec}
\end{figure}

To provide tighter constraints on the spectral behaviour of the afterglow the UVOT and XRT data were combined to produce a broadband spectrum. Good XRT events were extracted from time interval T+350 to T+950~s after the prompt emission, in order to exclude the X-ray flare. This only provided 27 events, which were therefore binned into one data point. Using Cash statistics, the best-fit parameters for a power law model with Galactic absorption and reddening and neutral gas incorporated at a free floating redshift were $\beta = -0.84^{+0.11}_{-0.06}$ and $\mathrm{z} = 2.44^{+0.30}_{-0.35}$, consistent with the results from a $\chi^2$ statistics fit, which had a reduced $\chi ^{2} = 0.6$ (2.3/3 d.o.f.). The derived redshift is typical of that measured in the {\it Swift} sample, which has a mean redshift of $\mathrm{z}=2.8$~\citep{jlf+05}, and the best-fit spectral index is in agreement with that inferred from the X-ray data alone ($\beta_{\mathrm{X}}=-1.1\pm 0.3$). There is therefore no indication of a spectral break between the X-ray and optical bands at this time. This fit is shown in Figure~\ref{fig:spec}.

\section{Discussion}\label{sec:disc}
The optical observations of the XRF~050406 taken with the {\it Swift}/UVOT telescope comprise the earliest detection of an X-ray Flash optical counterpart to date. Combining the optical/UV observations with simultaneous X-ray data provide multiwavelength temporal and spectral information on the characteristics of the very early burst afterglow. Analysis of this early time data constrains the nature of XRFs and the various theoretical models that have been put forward to explain their origin.

\subsection{Temporal information: a combined X-ray/optical analysis}
The flare observed in the X-ray light curve has been ascribed to emission from internal shocks several hundred seconds after the end of the $\gamma$-ray emission~\citep{rmb+05,brf+05}. Previous early observations made of GRB optical counterparts have shown evidence for both uncorrelated (GRB~990123; Akerlof et al. 1999, GRB~050401; Rykoff et al. 2005) and correlated (GRB~041219a; Vestrand et al. 2005, Blake et al. 2005) $\gamma$-ray and optical behaviour. However, the lack of such a flare in the optical/UV flux suggests that there is little contribution to the optical emission from internal shocks in this burst. Excluding the flare, the X-ray light curve is well fit by a broken power-law with a best-fit early time index $\alpha_{X,1} = -1.58\pm 0.17$, flattening out to a decay rate of $\alpha_{X,2} = -0.5^{+0.14}_{-0.13}$ at $\sim 4400$~s~\citep{rmb+05}. 

The X-ray decay rate before the break and the X-ray spectral index ($\beta_{\mathrm{X}}=-1.1\pm 0.3$) are consistent with afterglow emission in the fast-cooling regime for $\nu _{X} > \nu _{m}$, or the slow-cooling regime for $\nu _{X} > \nu _{c}$~\citep{rmb+05}, where an electron power-law distribution index p=2.5 is obtained in both cases. Assuming an electron power-law distribution index p=2.5, the optical decay index is only consistent with a fireball in the fast-cooling regime interacting with a wind circumburst environment, where $\nu_{\mathrm{opt}} < \nu_{c}$~\citep{zfd+05}. However, the flattening observed in the X-ray light curve cannot be explained with such a model.

The X-ray decay index after $\sim 4400$~s is consistent with that observed in the optical, and the flat decay in both energy bands could be the result of long lasting engine activity that continues injecting energy into the fireball through refreshed shocks~\citep{sm00,zm01}. Such a mechanism has been invoked to explain the shallow decay observed in GRB~050319~\citep{mas+05} and GRB~050401~\citep{dbb+05}. The continuous luminosity injection of the central engine is denoted by $L\propto t^{-q}$, and it is dominant over the fireball emission when $q < 1$~\citep{zm01}. The fireball model with additional energy injection is consistent with observations for $0 < q < 0.5$~\citep{zfd+05,rmb+05}, and a wind environment in the slow cooling regime gives temporal decay indices closest to observations.

A small spread in the Lorentz factor of the ejecta could also cause a shallow decay flux in the range of $-0.2\lesssim\alpha\lesssim -0.8$. Faster shells with larger Lorentz factors would be slowed down through interactions with the surrounding circumburst material, and as the slower shells caught up, on-going shocks would reduce the cooling rate~\citep{gk05}. This `shallow' flux decay would only be expected to continue until $\sim 10^4$~s, after which the Blandford-McKee self similar external shock evolution would become dominant and cause a steepening in the temporal decay, as is observed for a significant number of the {\it Swift} GRB sample~\citep{nkg+05}. In the case of XRF~050406 the X-ray light curve continues to decay at a constant rate with power law index $\alpha_{X,1} = -0.5^{+0.14}_{-0.13}$, until it is no longer detected after $\sim 10^6$~s. In order to account for this extended period of shallow flux additional factors have to be incorporated into this model~\citep{zfd+05} (see Romano et al. 2005). Another explanation is that the shallow decay is caused by a structured jet observed off-axis, where the amount of flux received from the wings of the jet would increase with time as the ejecta slowed down and the Lorentz factor decreased~\citep{pmg+05}. 

If the shallow decay is the result of continual engine activity or emission from late-time shocks, the early X-ray steep decay can only be attributed to the standard afterglow if the fireball energetics during this time dominate. Alternatively, the steep decline could be the result of an additional flux component no longer significant after $\sim 4400$~s. This source of hard emission could be either the tail of the prompt emission or continual X-ray flaring, both of which would not be expected to contaminate the optical band. Such flaring has been observed in other {\it Swift} GRBs~\citep{tgc+05}, although they typically have decay indices in the range $3\lesssim\alpha_{X,1}\lesssim 5$. The curvature effect should cause the light curve to fall as $\alpha = -2 -\beta$~\citep{kp00}, which is much steeper than the decay observed in the X-ray before $\sim 4400$~s. However, this decay rate assumes a uniform jet that is powered by a central engine which abruptly turns off at some time, after which the high-latitude emission is observed. A model in which the central engine activity is reduced would produce a shallower decay, as would also an off-axis structured jet, where both the energy and Lorentz factor vary with jet angle~\citep{dzf05}. At viewing angles larger than the core jet angle, emission from the core would be observed at later times, and the drop of flux would therefore be slower. An additionally factor that can also flattern the tail emission decay is an underlying forward shock component.

\subsection{Optical/UV observations in the context of XRF models}
\subsubsection{High-z GRB}
The detection in the three optical filters and upper limits obtained at wavelengths blue-ward of $\sim 4000$~\AA~firmly constrain the redshift of the burst to $\mathrm{z} < 3.1$, beyond which the Lyman edge is shifted into the optical wavebands. A model fit to the broadband spectrum that incorporates a Lyman absorption edge provides a best-fit redshift of $\mathrm{z}=2.44^{+0.30}_{-0.35}$. At such a redshift XRF~050406 would have an $E^{\mathrm{rest}}_{\mathrm{p}}\sim 52$~keV, nearly a factor of 10 lower than that of a GRB at $\mathrm{z}\sim 1$ with a typical observed peak energy value $E_{\mathrm{p}}=255$~keV. The softer spectrum observed in XRF~050406 cannot, therefore, be solely accounted for by a high-redshift burst. 

\subsubsection{Baryon loaded fireball}
A decelerating baryon loaded fireball ($\Gamma_{0}\ll 300$), referred to as a `dirty' fireball, can produce transient emissions that are longer lasting and most luminous at X-ray energies~\citep{dcb99}. In such a model the non-thermal synchrotron emission is produced by a relativistic blast wave that decelerates and radiates by sweeping up particles from the circumburst medium (external shock model), and the hardness and duration of the burst is a function of the baryon load. The observed peak frequency is strongly dependent on $\Gamma$, with $\nu_{m}\propto\Gamma^{4}$~\citep{mrw98} and the timescale over which the bulk of the power is radiated is the deceleration timescale $t_{d}$~\citep{rm92} where $t_{d}\propto \Gamma^{-8/3}$. Thus a dirty fireball with an initial Lorentz factor $\Gamma_{0}\leq 50$ makes the blast wave very inefficient at emitting $\gamma$-rays, and produces a long lasting soft burst.

A soft burst with a peak energy value $E_{\mathrm{p}}$ of e.g. an order of magnitude smaller than that observed in typical GRBs ($E_{\mathrm{p}}\sim 300$~keV) would have a $\Gamma$ value a factor of around two smaller. Such a burst would therefore produce a burst with peak power output lasting a factor of $\sim 6$ longer. The prompt emission from XRF~050406 only lasted 5.7~s, which is around a factor of 6 shorter than typical long GRBs, whose duration distribution peaks at $\sim 35$~s. Furthermore, at a redshift of 2.44 the rest frame burst duration would be $\sim 1.7$~s, placing it into the extreme short tail of long GRBs. A dirty fireball model therefore fails to simultaneously account for the shortness of this burst as well as its softness.

\subsubsection{Clean fireball}
An alternative mechanism that can produce a burst with soft prompt emission is an internal shock model in which the fireball has a high average Lorentz factor $\bar{\Gamma}$, and a fairly even $\Gamma$ distribution within the ejecta~\citep{zm02,bdm+05}. Using a simplified internal shock model to simulate a large number of bursts \citet{bdm+05} found that ejected matter with an overall large average Lorentz factor $\bar{\Gamma}$ but small variation in the Lorentz factor within the material reproduced well the observed properties of XRFs. Whereas the dependence on redshift, duration and injected power remained fairly similar between XRFs and GRBs, the small contrast in $\Gamma$ between colliding shells reduced the efficiency of energy dissipation through internal shocks, thus causing the prompt emission of these bursts to be weak and soft. A measure of the efficiency of a burst is provided by the ratio of the $\gamma$-ray fluence in the 15-350~keV energy band, $S_{\gamma}$, to the X-ray flux during its afterglow phase, 1 hour after the burst~\citep{lz04}. For typical GRBs this ratio produces a continuum on a log~$S_{\gamma}$-log~$\mathrm{F}_{X,1\mathrm{hr}}$ plot~\citep{rsf+05}. If the softer spectrum observed in XRF~050406 is indeed the result of lower efficiency during the prompt phase of the burst then one would expect the $S_{\gamma}/\mathrm{F}_{X,1\mathrm{hr}}$ ratio to be small relative to classical GRBs. On a log~$S_{\gamma}$-log~$\mathrm{F}_{X,1\mathrm{hr}}$ plot XRF~050406 lies within a $2\sigma$ band of the $S_{\gamma}/\mathrm{F}_{X,1\mathrm{hr}}$ distribution for a sample of 22 GRBs (see Fig.~3 of Roming et al. 2005b). There is therefore no evidence for lower efficiency in the prompt emission of XRF~050406. 

\subsubsection{Off-axis GRB}
Another possible model to explain the observed properties of XRFs is a GRB viewed off-axis. There are two main sub-categories related to the structure of the collimated flow: a uniform jet or a structured jet. In a uniform jet the Lorentz factor and energy distribution is constant within the jet but has a sharp cutoff at the edges, whereas in the structured jet model the bulk Lorentz factor and the energy per unit solid angle vary with the angular distance from the jet-axis $\theta$.

In the case of a uniform jet an XRF is presumed to be a GRB observed at an angle greater than the jet half-opening angle, $\theta_{v}>\theta_{j}$. Radiation is beamed into an angle $\theta = 1/\Gamma < \theta_{j}$, where $\Gamma$ is the initial bulk Lorentz factor of the relativistic outflow, and at lines of sight outside of the jet, the bulk of the radiation that is observed is at lower energies with frequencies $\nu\sim \nu'/\delta$, where $\delta\equiv\Gamma[1-\beta(\theta_{v}-\theta_{j})]\simeq [1+\Gamma^{2}(\theta_{v}-\theta_{j})^{2}]/2\Gamma$ is the Doppler factor~\citep{wl99,yin02}; $\nu$ therefore decreases as the viewing angle increases. As the Lorentz factor drops the radiation is beamed into a larger angle and a greater portion of the radiation reaches the observer. This, combined with the fading nature of the source afterglow, produces a flatter or even rising early-time light curve relative to that observed from a GRB viewed on-axis~\citep{gpk+02,rlr02}, and it peaks when $\Gamma\sim \theta_{v}^{-1}$. From here on the radiation from the jet core is beamed into the line of sight, and the light curve then begins to decay.

For on-axis GRBs a break in the light curve is expected at a time $t_{\mathrm{break}}\propto\theta_{j}^{8/3}$~\citep{sph99}, where $\theta_{j}$ is the jet opening angle, and this is typically ascribed to breaks observed on the order of several hours to days after the prompt emission. In the case of a GRB observed off-axis, $\theta_{v} > \theta_{j}$, and therefore a break in the light curve occurs at a later time than is typically observed for on-axis GRBs. Assuming a jet opening angle of $\theta_{j}=1^{o}$, \citet{rlr02} modelled the resultant afterglow light curve observed from a GRB viewed at various off-axis angles, $\theta_{v}$. They found a viewing angle $\theta_{v}=1^{o}$, close to the edge of the jet, produced a peak in the light curve at t $> 1000$~s. Any peak in the optical light curve of XRF~050406 occurred in the first few hundred seconds after the prompt $\gamma$-ray emission. In an off-axis uniform jet model this would require an anomalously small jet opening angle, which causes difficulties in standard afterglow models and observations~\citep{ldz04,zdl+04}.

Alternatively we consider a structured jet in which the energy $\epsilon$ and Lorentz factor $\Gamma$ per unit solid angle is a decreasing function of angular distance from the centre $\theta$. In such a model a softer spectrum can be observed for viewing angles $\theta_{v}<\theta_{j}$, provided that they are close to the outer edges of the jet where the energy per solid unit angle is lower. From the onset the emission from material in the observer's line of sight will dominate over that produced within the jet core. The light curve is, therefore, primarily composed of a single decaying component from emission regions in the line of sight, with an underlying flux component originating from the core of the jet. The resultant light curve would be observed to decay from the start but at a slower rate than if observed on axis. This model is in good agreement with observations of XRF~050406, which has a shallow optical decay index $\alpha_{opt} = -0.75\pm 0.26$, and late time shallow X-ray decay index $\alpha_{X,2} = -0.5^{+0.14}_{-0.13}$.

Another consequence of a structured jet model is a smoother prompt light curve. As in the case of the afterglow, the prompt emission will be dominated by material along the line of sight. At large viewing angles strong variability in the Lorentz factor is hard to produce~\citep{rl02, zwh04}, causing internal shocks to be an inefficient source of radiation. External shocks are therefore expected to dominate the prompt emission and this produces a much smoother prompt light curve containing a single wide peak. This is indeed what is observed in the case of XRF~050406.

Within the universal structured jet model several jet morphologies have been explored with varying functional dependencies on the energy density and Lorentz factor on the jet axis angle. A simple power-law structure, in which the energy varies with jet axis angle as $\theta^{-2}$, has difficulties reproducing the observed XRF, XRR burst, and GRB distribution~\citep{ldg05,zdl+04}, and greatly over-predicts the number of XRFs. \citet{zdl+04} addressed this by proposing a Gaussian-like jet energy distribution. Several other models have been explored, such as a jet with two or more components~\citep{bkp+03,sfw+03,hwd+04}. However a much greater data sample of bursts with early time observations is required before the range of possible jet morphologies can be narrowed down. 

It is important to note that an off-axis GRB should appear less luminous than one observed on-axis as a result of a lower Lorentz factor in the emission regions of the observed radiation. This is consistent with both optical and X-ray observations of XRF~050406, which had optical and X-ray fluxes in the order of $7\times 10^{-17}$~\flux ~and $10^{-10}$~\flux ~at $\sim 100$~s after the prompt burst emission. As a reference, this is at least an order of magnitude smaller than the bright burst GRB~050525 at similar timescales after the prompt emission. \citet{yin02,yin03} have argued that an XRF's lower luminosity should limit their detectability to only modest redshifts (z$<\sim 1$), so raising a question as to whether the best-fit redshift for XRF~050406 of $\mathrm{z}=2.44^{+0.30}_{-0.35}$ is consistent with it being a GRB viewed off-axis. The apparent paradox between the suggested off-axis viewing angle and redshift of XRF~050406 could be the cause of an intrinsically very luminous XRF, although the results of \citet{yin02} were reached on the basis of a GRB having a uniform, rather than a structured jet. Furthermore, with only small changes in the parameters of their model \citet{yin04} were able to extend the detectability of off-axis GRBs to redshifts of up to $\mathrm{z}\leq 4$. These included a decrease in the collimated jet angle down to angles $\lesssim 0.03$~rad, an increase in the assumed standard energy constant $E_{\gamma}$ from $0.5\times 10^{51}$~\flu ~to $1.15\times 10^{51}$~\flu, a steeper high energy spectral index from -2.5 to -3, and a modification of the spectral functional form of $f(\nu^{'})$.

\subsection{The Amati Relation}
If XRFs do indeed have the same origin as GRBs and XRR bursts they should follow the same empirical laws. It has been shown that the relation $E_{\mathrm{p}}\propto E_{\mathrm{iso}}^{1/2}$~\citep{aft+02}, applicable to GRBs, is also consistent with the $E_{\mathrm{p}}$ limits for XRF~020903 and XRF~030723~\citep{stk+03,ldg05}, and recently XRF~050416a has also been shown to satisfy the relation~\citep{sbb+05}. At the best-fit redshift value of $\mathrm{z}=2.44$, XRF~050406 has an equivalent isotropic energy of $E_{iso}=1.55\times 10^{51}$~erg, which when using the Amati relation, provides an estimated peak energy of $E_{\mathrm{p},\mathrm{Amati}}\approx 11.4$~keV in the observer frame. Although the $E_{\mathrm{p}}$ values from spectral fits to the prompt emission are not well constrained, they are consistent within errors to the estimated $E_{\mathrm{p,Amati}}$. This would give support to the hypothesis that GRBs and XRFs form a continuum, and that a unified model is therefore necessary to explain both classes of bursts.

\section{Conclusions}\label{sec:concs}
We have presented early-time optical data for XRF~050406, beginning 88~s after the burst trigger. This constitutes the earliest detection of an X-ray Flash in the optical band to date. The properties of the light curve and spectrum constrain the possible models that explain the soft spectrum from the prompt emission. The lack of Lyman absorption down to $4000$~\AA~in the UV/optical spectrum provides a firm upper limit of $\mathrm{z}\leq 3.1$ on the redshift, and broadband spectral modelling yields a best-fit redshift $\mathrm{z} = 2.44^{+0.30}_{-0.35}$, and spectral index $\beta = -0.84^{+0.11}_{-0.06}$. A fit to the optical light curve provides a decay rate of $\alpha = -0.75\pm 0.26$. Combining this information with the prompt $\gamma$-ray emission and X-ray afterglow we conclude that the soft spectrum observed in the initial emission from XRF~050406 is most likely the result of a GRB with a structured jet observed off-axis. This model is able to account for the lack of a peak in the X-ray or optical afterglow in the order of $10^3$~s or later after the prompt emission, as well as the shallow decay rate observed in both energy bands. It also predicts a smooth prompt emission light curve. 

In order to continue to further constrain the possible models that describe the origins of XRFs, further prompt observations of these objects are necessary. It is only during the early stages of emission, when the internal mechanisms that power these bursts are probed, that many of these models can be differentiated. It is also important to increase the sample of XRFs with spectroscopic redshifts to gain information on their rest-frame properties. The classification of bursts as X-ray Flashes is based on observer frame properties. However, only knowledge of the rest-frame properties will reveal the underlying differences between XRFs, XRR bursts and GRBs. At a redshift of $\mathrm{z}=2.44$ XRF~050406 would have a peak energy $E_{\mathrm{p}}\sim 52$~keV, at which stage there begins to be an overlap between the intrinsic properties of XRFs that are at redshifts $\mathrm{z} > 1-2$ and nearby GRBs. Continued prompt and deep observations of XRFs attainable with {\it Swift} together with an increase in spectroscopic redshifts acquired from rapid follow-up ground-based observations promise to provide the data necessary to further constrain the possible origins of XRFs.

\acknowledgments
This work is sponsored at Penn State by NASA contract NAS5-00136; at OAB by funding from ASI on grant number I/R/039/04; and at UCL-MSSL and the University of Leicester by PPARC.

We gratefully acknowledge the contributions of all members of the {\it Swift} team. Patricia Schady acknowledges the support of a PPARC quota studentship.



\clearpage

\clearpage

\begin{deluxetable}{cccll}
\tabletypesize{\scriptsize}
\tablewidth{0pt}
\tablecaption{UVOT Time Resolved Photometry\label{tab:timephot}}
\tablehead{
\colhead{Filter} & \colhead{Mid Time (s)} & \colhead{Exposure (s)} & \colhead{Counts/s} & \colhead{Magnitude}}
\startdata
{\it V}\tablenotemark{a} & 113 & 50 & $0.38\pm 0.12$ & $18.92\pm 0.31$ \\
{\it V}\tablenotemark{a} & 163 & 50 & $0.18\pm 0.10$ & $19.76\pm 0.59$ \\
{\it V}\tablenotemark{a} & 105 & 35 & $0.17\pm 0.13$ & $19.79\pm 0.73$ \\
{\it V}\tablenotemark{a} & 140 & 35 & $0.32\pm 0.14$ & $19.11\pm 0.43$ \\
{\it V}\tablenotemark{a} & 173 & 30 & $0.36\pm 0.15$ & $18.99\pm 0.42$ \\
{\it V}\tablenotemark{b} & 138 & 100 & $0.25\pm 0.08$ & $19.44\pm 0.32$ \\
{\it V}\tablenotemark{b} & 598 & 90 & $< 0.09$ & $> 20.51$ \\
{\it V}\tablenotemark{b} & 11011 & 900 & $< 0.03$ & $> 21.59$ \\
{\it B} & 240 & 10 & $0.32\pm 0.42$ & $20.40\pm 1.33$ \\
{\it B} & 329 & 10 & $< 0.36$ & $> 20.27$ \\
{\it B} & 423 & 10 & $0.72\pm 0.47$ & $19.52\pm 0.66$ \\
{\it B} & 517 & 10 & $0.78\pm 0.47$ & $19.43\pm 0.61$ \\
{\it B} & 641 & 10 & $1.06\pm 0.50$ & $19.10\pm 0.48$ \\
{\it B} & 706 & 10 & $0.38\pm 0.41$ & $20.21\pm 1.08$ \\
{\it B} & 801 & 10 & $0.59\pm 0.43$ & $19.73\pm 0.73$ \\
{\it B} & 891 & 10 & $0.15\pm 0.34$ & $21.22\pm 2.23$ \\
{\it B} & 990 & 10 & $< 0.26$ & $> 20.62$ \\
{\it B}& 6029 & 900 & $< 0.04$ & $> 22.67$ \\
{\it U} & 226 & 10 & $0.68\pm 0.40$ & $18.80\pm 0.59$ \\
{\it U} & 314 & 10 & $0.30\pm 0.32$ & $19.69\pm 1.06$ \\
{\it U} & 407 & 10 & $0.85\pm 0.43$ & $18.56\pm 0.50$ \\
{\it U} & 503 & 10 & $0.75\pm 0.41$ & $18.69\pm 0.55$ \\
{\it U} & 595 & 10 & $0.13\pm 0.27$ & $20.60\pm 2.01$ \\
{\it U} & 691 & 10 & $0.37\pm 0.33$ & $19.46\pm 0.91$ \\
{\it U} & 785 & 10 & $0.25\pm 0.28$ & $19.90\pm 1.11$ \\
{\it U} & 879 & 10 & $0.11\pm 0.25$ & $20.78\pm 2.33$ \\
{\it U} & 973 & 10 & $< 0.20$ & $> 20.13$ \\
{\it U} & 5221 & 900 & $< 0.04$ & $> 21.82$ \\
{\it UB}\tablenotemark{c} & 233 & 20 & $0.43\pm 0.23$ & $19.75\pm 0.53$ \\
{\it UB} & 322 & 20 & $0.28\pm 0.22$ & $20.22\pm 0.79$ \\
{\it UB} & 415 & 20 & $0.69\pm 0.26$ & $19.24\pm 0.38$ \\
{\it UB} & 510 & 20 & $0.64\pm 0.28$ & $19.32\pm 0.44$ \\
{\it UB} & 618 & 20 & $0.48\pm 0.25$ & $19.64\pm 0.52$ \\
{\it UB} & 699 & 20 & $0.27\pm 0.25$ & $20.26\pm 0.93$ \\
{\it UB} & 793 & 20 & $0.35\pm 0.21$ & $19.98\pm 0.60$ \\
{\it UB} & 885 & 20 & $0.39\pm 0.23$ & $19.86\pm 0.59$ \\
{\it UB} & 982 & 20 & $< 0.17$ & $> 20.74$ \\
{\it UB} & 5625 & 1800 & $< 0.03$ & $> 22.73$ \\
\enddata
\tablenotetext{a}{Event mode data.}
\tablenotetext{b}{Image mode data.}
\tablenotetext{c}{A {\it UB} filter zero-point of 18.84 is assumed. This is the magnitude of an object that produces one count per second in a combined UB filter given that a zero magnitude object has a count rate of 1/2(cr(B)+cr(U)). cr(B) and cr(U) are the B and U band count rates of a zero magnitude object, correspondingly.} 
\end{deluxetable}

\begin{deluxetable}{ccclc}
\tablewidth{0pt}
\tablecaption{UVOT Spectral Photometry\label{tab:specphot}}
\tablehead{
\colhead{Filter} & \colhead{Effective Wavelength} & \colhead{Exposure} & \colhead{Counts/s} & \colhead{Flux}\\
& \colhead{(\AA)} & \colhead{(s)} & & \colhead{($10^{-16}$~\nuflux)}}
\startdata
{\it V} & 5430 & 64 & $0.119\pm 0.108$ & $0.266\pm 0.241$\\
{\it B} & 4340 & 76 & $0.405\pm 0.126$ & $0.532\pm 0.165$\\
{\it U} & 3440 & 73 & $0.505\pm 0.114$ & $0.751\pm 0.170$\\
{\it UVW1} & 2600 & 73 & $0.003\pm 0.024$ & $0.009\pm 0.084$\\
{\it UVM2} & 2200 & 71 & $0.011\pm 0.014$ & $0.075\pm 0.097$\\
{\it UVW2} & 1930 & 63 & $0.035\pm 0.028$ & $0.214\pm 0.166$\\
\enddata
\end{deluxetable}

\clearpage

\clearpage

\clearpage

\end{document}